\documentclass[a4paper,11pt,reqno]{amsart}
\usepackage[a4paper]{geometry}
\usepackage{amssymb,amsmath}
\usepackage{mathtools}
\usepackage{enumerate}
\usepackage{bookmark}
\usepackage{hyperref}
\usepackage[initials,backrefs]{amsrefs}

\numberwithin{equation}{section}
\theoremstyle{plain}
\newtheorem{theorem}{Theorem}
\newtheorem*{assi*}{(H.1) Short-range interaction}
\newtheorem*{assp*}{(H.2) External potential}
\newtheorem*{assr*}{(H.3) Bounded resolvent}
\newcommand{\prob}[1]{\DP\left\{#1\right\}}
\newcommand{\Bone}{\mathbf{1}}
\newcommand{\Bzero}{\mathbf{0}}
\newcommand{\DP}{\mathbb{P}}
\newcommand{\DR}{\mathbb{R}}
\newcommand{\DZ}{\mathbb{Z}}
\newcommand{\DN}{\mathbb{N}}
\newcommand{\BC}{\mathbb{C}}
\newcommand{\BDelta}{\mathbf{\Delta}}
\newcommand{\Bx}{\mathbf{x}}
\newcommand{\By}{\mathbf{y}}
\newcommand{\BH}{\mathbf{H}}
\newcommand{\BU}{\mathbf{U}}
\newcommand{\BV}{\mathbf{V}}
\newcommand{\FB}{\mathfrak{B}}
\newcommand{\supp}{\mathrm{supp}}
\newcommand{\condI}{\mathbf{(H.1)}}
\newcommand{\condP}{\mathbf{(H.2)}}
\newcommand{\condR}{\mathbf{(H.3)}}
\newcommand{\tto}[1]{\smash{\mathop{\,\,\,\, \longrightarrow \,\,\,\, }\limits_{#1}}}

\begin{document}
\title[Multi-particle Lifshitz tails]{Lifshitz tails for the multi-particle continuous Anderson model}

\author[T.~Ekanga]{Tr\'esor EKANGA$^{\ast}$}

\address{$^{\ast}$%
Institut de Math\'ematiques de Jussieu,
Universit\'e Paris Diderot,
Batiment Sophie Germain,
13 rue Albert Einstein,
75013 Paris,
France}
\email{tresor.ekanga@imj-prg.fr}

\begin{abstract}
We consider the multi-particle Anderson model in the continuum and show that under some mild assumptions on the inter-particle interaction and the external potential, its lower spectral edge is almost surely constant and is the same with that of the single-particle model. We then obtain the lifshitz asymptotics for the multi-particle hamiltonian in the continuum near the bottom of the spectrum.
\end{abstract}
\maketitle

\section{Introduction}
The study of multi-particle random Schr\"odinger operators is a quite recent new direction in the mathematics of random Scr\"odinger operators. In the single-particle theory, a lot of results on localization, continuity of the integrated density of states and Lifshitz asymptotics have been obtained for different forms of the external potentials see for example \cites{BCH97,CKM87,DK89, CL90,St01,K08,PF92,RS78} for the Anderson localization, \cites{CHK03} for continuity of the integrated density states and \cites{KP99,KS86} for the Lifshitz tails. See also the references therein.

For multi-particle systems, we have to distinguish the lattice case with the continuum case. First of all, let us say that today, there is not yet a proof of the existence of the integrated density of states for multi-particle systems on the lattice. In the continuum case, the works by Klopp and Zenk \cites{KZ03,Z03} for multi-particle homogeneous models establish under some general assumptions, the existence of the integrated density states. Actually, their result is more precised, they showed that the interacting multi-particle integrated density of states exists and is the same with the single-particle one. The proof of Klopp and Zenk uses the Helffer-Sj\"ostrand formula of the functional calculus.

Although, the object of the paper is not localization, for the reader convenience, we mention some works on Anderson localization for multi-particle systems in both the discrete and the continuum cases \cites{AW09,AW10,BCSS10,BCS11,CS09b,C11,C12,E11,E12,E13,E16,FW15,KN13}.

In this paper, we analyze the bottom of the spectrum of the multi-particle continuous Anderson model under fairly general assumptions on the inter-particle interaction and the random external potential. In fact, proving that the multi-particle lower spectral edge is the same with the single-particle one is the heart of the paper. Note, that in our previous works \cites{E11,E12} for multi-particle systems at low energy on the lattice, we studied the bottom of the spectrum in  order to show that it is non-random in absence of ergodicity. We use the general concept and ideas of \cite{E12} and adapt them in the continuum case.

Let us now discuss on the results and the structure of the paper. Our main results for multi-particle systems in the continuum are Theorem \ref{thm:lower.spectral.edges} (the multi-particle lower spectral edges are non-random) and Theorem  \ref{thm:LT.mp} (multi-particle Lifshitz tails). In the rest of this section, we describe the multi-particle Anderson model in the continuum and  the main assumptions.  Sections \ref{sec:lower.spectral.edges} and \ref{sec:LT.mp} are devoted to the statements of the results. Finally, in section \ref{sec:proofs.results}, we prove the results. 

\subsection{The model }

We fix at the very beginning the number of particles $N\geq 2$. We are concern with multi-particle random Schr\"odinger operators of the following forms:
\[
\BH^{(N)}(\omega):=-\BDelta + \BU+\BV,
\]
acting in $L^{2}((\DR^{d})^N)$. Sometimes, we will use the identification $(\DR^{d})^N\cong \DR^{Nd}$. Above, $\BDelta$ is Laplacian on $\DR^{Nd}$, $\BU$ represents the inter-particle interaction which acts as multiplication operator in $L^{2}(\DR^{Nd})$. Additional information on $\BU$ is given in the assumptions. $\BV$ is the multi-particle random external potential also acting as multiplication operator on $L^{2}(\DR^{Nd})$. For $\Bx=(x_1,\ldots,x_N)\in(\DR^{d})^N$, $\BV(\Bx)=V(x_1)+\cdots+ V(x_N)$ and $\{V(x,\omega), x\in\DR^d\}$ is a random i.i.d. stochastic process relative to some probability space $(\Omega,\FB,\DP)$.

Observe that the non-interacting Hamiltonian $\BH^{(N)}_0(\omega)$ can be written as a tensor product:
\[
\BH^{(N)}_0(\omega):=-\BDelta +\BV=\sum_{k=1}^N \Bone^{\otimes(k-1)}_{L^{2}(\DR^d)}\otimes H^{(1)}(\omega)\otimes \Bone^{\otimes(N-k)}_{L^2(\DR^d)},
\]
where, $H^{(1)}(\omega)=-\Delta + V(x,\omega)$ acting on $L^2(\DR^d)$. We will also consider random Hamiltonian $\BH^{(n)}(\omega)$, $n=1,\ldots,N$ defined similarly. Denote by $|\cdot|$ the max-norm in $\DR^{nd}$.

\subsection{The assumptions}
We now describe our general assumptions on the continuous multi-particle random Hamiltonian.

\begin{assi*}
The global interaction $\BU$ is of the form:
\[
\BU(\Bx)=\sum_{1\leq i<j\leq N} U(|x_i-x_j|),
\]
where the function $U: \DR\rightarrow\DR$ is square integrable and non-negative. Further it is plain that $U$ is of finite range, i.e., $\exists r_0>0$ such that for $x,y\in\DR^d$  with $|x-y|>r_0$, we have $U(|x-y|)=0$.
\end{assi*}

For the external potential $V$, we assume:

\begin{assp*}
The potential $V$ is square integrable and non-negative. 
\end{assp*}
We have one more assumption
\begin{assr*}
The operator $\BU(\BH^{(N)}_0-i)^{-1}$ is bounded.
\end{assr*}
This last assumption  was essential in \cite{KZ03} and is valid for example in the particular case of the Coulomb and Yukawa potentials see \cite{RS78}.

\subsection{The results on the bottom of the spectrum }\label{sec:lower.spectral.edges}

For any $1\leq n\leq N$, we denote by $\sigma(\BH^{(n)}(\omega))$ the spectrum of $\BH^{(n)}(\omega)$ and by $E^{(n)}_0(\omega)$ the infimum of $\sigma(\BH^{(n)}(\omega))$. The main result of this subsection is 

\begin{theorem}[The multi-particle lower spectral edges are non-random]\label{thm:lower.spectral.edges}
Let $1\leq n\leq N$. Assume that the assumptions $\condI$, $\condP$ and $\condR$ hold true. Then with $\DP$-probability $1$, 

\[
0\in\sigma(\BH^{(N)}(\omega))\subset[0;+\infty).
\]
Consequently, for $n=1,\ldots,N$, $E^{(n)}_0=0$ almost surely.
\end{theorem}

\subsection{The result on the integrated density of states}\label{sec:LT.mp}
Let $1\leq n\leq N$ and $L>0$.  Denote by $\BC^{(n)}_L(\Bzero)=\{x\in(\DR^d)^n: |\Bx|<L\}$ the open cube on $(\DR^d)^n$. and $\BH^{(n)}_{\BC^{(n)}_L(\Bzero)}$ the restriction of $\BH^{(n)}(\omega)$ on the cube $\BC^{(n)}_L(\Bzero)$ with Dirichlet boundary conditions. We have

\begin{theorem}[Lifshitz tails]\label{thm:LT.mp}
Let $1\leq n\leq N$. Under assumptions $\condI$, $\condP$ and $\condR$, we have that for any $E\in \DR$, the limit
\[
\lim_{L\to\infty} L^{-nd}Trace(\Bone_{(-\infty;E]}(\BH^{(n)}_{\BC^{(n)}_L(\Bzero)})),
\]
exists and is denoted by $N(\BH^{(n)},E)$. Further, the quantity $N(\BH^{(n)},E)$ (called, the integrated density of  states of $\BH^{(n)}(\omega)$) satisfies the Lifshitz tails: there exist constants $C>0$ and $\gamma>0$ such that

\[
N(\BH^{(n)},E) \sim C\cdot\exp(-\gamma(E-E^{(n)}_0)^{-\frac{d}{2}})\quad \text{ as $E\searrow E^{(n)}_0$}.
\]
\end{theorem}

\section{Proof of the results}\label{sec:proofs.results}

\subsection{Proof of Theorem \ref{thm:lower.spectral.edges}}
Let $1\leq n\leq N$. We aim to prove that $0\in\Sigma(\BH^{(n)}(\omega)\subset[0;+\infty)$. Assumption $\condI$ implies that $\BU$ is non-negative and assumption $\condP$ also implies that $\BV$ is non-negative. Since, $-\BDelta\geq 0$, we get that almost surely $\sigma(\BH^{(n)}(\omega))\subset [0;+\infty)$. It remains to see that $0\in \sigma(\BH^{(n)}(\omega))$ almost surely. 

Let $k,m\in\DN$. Define,
\[
B_{k,m}:=\{\Bx\in\DZ^{nd}: \min_{i\neq j}|x_i-x_j|>r_0+2km\},
\]
where $r_0>0$, is the range of the interaction $\BU$. We also define the following sequence in $\DZ^{nd}$,
\[
\Bx^{k,m}:=C_{k,m}(1,\ldots,nd),
\]
where $ C_{k,m}= r_0+ 2km+1$. Using the identification  $\DZ^{nd}\cong(\DZ^d)^n$, we can also write $\Bx^{k,m}=C_{k,m}(x_1^{k,m},\ldots,x_n^{k,m})$ with each $x_i^{k,m}\in\DZ^d$, $i=1,\ldots,n$. Obviously each term $\Bx^{k,m}$ of the sequence $(\Bx^{k,m})_{k,m}$ belongs to $B_{k,m}$. For $j=1,\ldots,n$ set
\[
H^{(1)}_j(\omega):=-\Delta+V(x_j,\omega).
\]
We have that almost surely $\sigma(H^{(1)}_j(\omega))=[0;+\infty)$, see for example \cite{St01}. So, if we set for $j=1,\ldots,n$, 
\[
\Omega_j:=\{\omega\in\Omega: \sigma(H^{(1)}_j(\omega))=[0;+\infty)\}.
\]
$\prob{\Omega_j}=1$ for all $j=1,\ldots,n$. Now, put
\[
\Omega_0:=\bigcap_{j=1}^n \Omega_j.
\]
We also have that $\prob{\Omega_0}=1$. Let $\omega\in \Omega_0$, for this $\omega$, we have that $0\in \sigma(H^{(1)}_j(\omega))$ for all $j=1,\ldots,n$ and by the Weyl criterion, there exist $n$ Weyl sequences $\{(\phi_j^m)_m: j=1,\ldots,n\}$ related to $0$ and each operator $H^{(1)}_j(\omega)$. By the density property of compactly supported functions $C^{\infty}_c(\DR^d)$ in $L^2(\DR^d)$, we can directly assume that each $\phi^m_j$ is  of compact support, i.e., $\supp \phi_j^m\subset C^{(1)}_{k_jm}(0)$ for some integer $k_j$ large enough. Set 
\[
k_0=\max_{j=1,\ldots,n}k_j,
\]
and put, $\Bx^{k_0,m}=(x_1^{k_0,m},\ldots, x_n^{k_0,m})\in B_{k_0,m}$. We translate each function $\phi_j^m$ to have support contained in the cube $C^{(1)}_{k_0m}(x_j^{k_0})$. Next, consider the sequence $(\Phi^m)_m$ defined by the tensor product,
\[
\phi^m:=\phi_1^m\otimes\cdots\otimes\phi_n^m.
\]
We have that $\supp \phi^m\subset\BC^{(n)}_{k_0m}(\Bx^{k_0,m})$ and we aim to show that, $(\phi^m)_m$ is a Weyl sequence for $\BH^{(n)}(\omega)$ and $0$. For any $\By\in\DR^{nd}$:
\[
|(\BH^{(n)}(\omega)\phi^m)(\By)|=|(\BH^{(n)}_0(\omega)\phi^m)(\By)|.
\]
Indeed, for the values of $\By$ inside the cube $\BC^{(n)}_{k_0m}(\Bx^{k_0,m})$ the interaction potential $\BU$ vanishes and for those values outside that cube $\phi^m$ equals zero too. Therefore,
\begin{align*}
\|\BH^{(n)}(\omega)\phi^m\|&\leq\|\BH^{(n)}_0(\omega)\phi^m\|\\
&\leq \prod_{j=1}^n\| H_j^{(1)}(\omega)\phi_j^m\| \tto{m\to +\infty} 0,
\end{align*}
because, for all $j=1,\ldots,n$, $\| H^{(1)}_j(\omega)\phi_j^m\| \rightarrow 0$ as $m\rightarrow +\infty$, since $\phi_j^m$ is a weyl sequence for $H^{(1)}_j(\omega)$ and $0$. This complete the proof.

\subsection{Proof of theorem \ref{thm:LT.mp}}\label{ssec:proof.LT.mp}
By Theorem \ref{thm:lower.spectral.edges}, we know that all the lower spectral edges of $\BH^{(n)}(\omega)$, $n=1,\ldots,n$ are almost surely equal to $0$. Now, by the work of Klopp and Zenk \cite{KZ03}, we know that the multi-particle integrated density of states of each $\BH^{(n)}(\omega)$ exists and is the same with that of the single-particle. For single-particle Anderson models the integrated density of states exists see\cite{PF92} and in addition it admits the Lifshitz asymptotics see \cite{St01} for exmaple. Finally, since all the lower spectral edges are equal to zero, we conclude that the multi-particle Anderson model in the continuum also admits the Lifshitz tails near its lower spectral edge, i.e., zero. This complete the proof.

\end{document}